# Anomalous Transverse Response and Multi-Field Ferrialtermagnetic-Ferroelectric Valve with CrSb Flakes

Long Zhang[1], Xinfeng Chen[2], Hongfei Liang[1], Jianting Dong[1], Fei Zou[1], Yi Yan[1], Guangqian Ding[3,*], Guoying Gao[1,*]

[1]School of Physics, Wuhan National High Magnetic Field Center, Hubei Fundamental Research Center for Physics, Huazhong University of Science and Technology, Wuhan, 430074 China

[2]Key Laboratory of Advanced Optoelectronic Quantum Architecture and Measurement (MOE), Beijing Key Laboratory of Nanophotonics and Ultrafine Optoelectronic Systems, and School of Physics, Beijing Institute of Technology, Beijing 100081, China

[3]School of Electronic Science and Engineering, Chongqing University of Posts and Telecommunications, Chongqing 400065, China

*Corresponding authors. Emails: dinggq@cqupt.edu.cn (G.D.); guoying_gao@mail.hust.edu.cn (G.G.)

**ABSTRACT:** Altermagnets combine the zero-stray-field of antiferromagnets with the spin polarization of ferromagnets, showing great potential for spintronic applications. Here, we propose ferrialtermagnetism as a distinct subclass of altermagnetic family, where symmetry-inequivalent altermagnetic sublattices possess nonidentical Néel vectors, preventing mutual cancellation of alternating spin splitting and conferring intrinsic robustness against perturbations. This concept is realized in the three-atomic-layer CrSb (110) flakes, which exhibits spin splitting of 344 meV, moderate uniaxial magnetic anisotropy, and high Néel temperature of 657 K. The magneto-optical Kerr and the anomalous Hall effects are observed. Integrating this ferrialtermagnetic CrSb with ferroelectric $Sc_2CO_2$ and Cu spacer, we design an ferrialtermagnetic-ferroelectric valve. This device displays equilibrium tunneling magnetoresistance and electroresistance of ~$10^3$%, and non-equilibrium magnitudes under bias, thermal, or light field reaches ~$10^4$% with high spin filtering of 90%. The negative differential resistance and photogalvanic effects, and photocurrent extinction ratio of 283.8 are achieved. These findings establish ferrialtermagnetism as a fertile platform for multi-field-controlled, ultracompact, and self-powered spintronics and electronics.

## INTRODUCTION

The recent identification of altermagnets has established a distinct collinear magnetic phase that combines the advantages of conventional antiferromagnets and ferromagnets (*1-3*). Altermagnetic (AM) order is characterized by two sublattices related by rotation (*C*) or mirror (*M*) symmetry. This coupling breaks time-reversal symmetry while preserving vanishing net magnetization (*4, 5*). Consequently, a momentum-dependent non-relativistic spin splitting emerges, enabling ferromagnet-like phenomena including the magneto-optical Kerr effect (MOKE) and the anomalous Hall effect (AHE) (*6, 7*). Exemplary AM compounds including CrSb(*8*), MnTe(*9*), $KV_2Se_2O$ (*10*), $RuO_2$ (*11*), etc., have been synthesized and widely investigated (*12-15*). Their compelling attributes, including zero stray field, ultrafast spin dynamics, low power consumption, and high integration density, position altermagnets as promising candidates for next-generation spintronic and electronic devices. Moreover, the integration of altermagnetism with ferroelectric (FE) order simultaneously breaks both time-reversal and inversion (*P*) symmetries and opens a pathway toward multistate control of charge and spin transport (*16-20*).

Conventional antiferromagnetic (AFM) systems (Fig. 1a), in contrast to ferromagnetic (FM) ones, preserve time-reversal symmetry. Their opposing magnetic sublattices are connected via *P* or translation ($\tau$) operations, enforcing complete spin degeneracy and zero net magnetization. Conversely, ferrimagnetic (FiM) order lacks such symmetry mapping between sub-lattices, yielding intrinsic spin splitting. The recently proposed antialtermagnetic (AAM) state (*21*), also termed hidden altermagnetism (*22*), consists of AM sublattice pairs (each comprising two opposite sublattices) that are connected by $P$ or $\tau$ symmetry, leading to an overall spin-degenerate band structure despite internal alternating spin splitting. Extending this symmetry-based classification, we propose a distinct subclass: ferrialtermagnetism (Fig. 1a). In a ferrialtermagnetic (FiAM) system, the constituent AM sublattices are symmetry-inequivalent and exhibit Néel vectors that differ in magnitude. Notably, the alternating spin splitting induced by distinct sublattice pairs is nonidentical, precluding mutual cancellation when the internal Néel vectors of a subset of AM sublattice pairs are reversal. This intrinsic asymmetry confers robustness against external perturbations.

Realizing this FiAM state necessitates a suitable material platform. Bulk NiAs-type CrSb exhibits a Néel temperature ($T_N$) of ~700 K (*23*) and definitive AM spin splitting (*8*). Two-dimensional (2D) variants with varying facets adopting the NiAs structure have been extensively synthesized (*24, 25*). Building on our recent investigation of phase-, facet-, and thickness-dependent altermagnetism in CrSb (*26*), we focus here on the three-atomic-layer CrSb (110) facet. In this geometry, the inequivalence between the inner atomic layer and the two surface layers precludes any symmetry operation mapping the inner Cr pairs onto the outer ones. This symmetry breaking directly corresponds to the FiAM state proposed above. Elucidating this previously unrecognized subtype of altermagnetism, and exploring its synergy with multiferroic control, promises to significantly enrich the AM family and broaden its technological scope.

In this work, we investigate three-atomic-layer CrSb (110) flakes as a FiAM representative via symmetry analyses, density functional theory (DFT), Monte Carlo (MC) simulations, Wannier function, and non-equilibrium Green's function (NEGF) approaches. The FiAM CrSb shows alternating spin splitting of 344 meV, moderate uniaxial magnetic anisotropy, a high $T_N$ of 657 K, MOKE, and AHE. We propose an AM-FE valve constructed by $CrSb/Sc_2CO_2/Cu/Sc_2CO_2/CrSb$, where tunneling magnetoresistance (TMR) and tunneling electroresistance (TER) reach equilibrium ~$10^3$% and non-equilibrium ~$10^4$% with high spin filtering efficiencies around 90%. The negative differential resistance (NDR) effect under bias, photogalvanic effect, and several dozen to a few hundred extinction ratios (ERs) can be achieved. These results open avenues for multi-field miniaturized multiferroic transport applications.

## RESULTS

### Structure and Ground-State Altermagnetism

The three-atomic-layer 2D NiAs-type CrSb flakes, stoichiometrically $Cr_6Sb_6$ but denoted CrSb for brevity, possesses symmetry operations including $m_x\tau_{(1/2\ 0\ 0)}$, $m_y$, $m_z\tau_{(1/2\ 0\ 0)}$, $C_{2x}\tau_{(1/2\ 0\ 0)}$, $C_{2y}$, $C_{2z}\tau_{(1/2\ 0\ 0)}$, identity ($I$), and $P$. Its magnetic ground state is AFM1 (Fig. 2a,b). Scanning tunneling microscopy (STM) reveals pronounced charge accumulation around the surface Cr and Sb atoms (Fig. 3a), suggesting favorable electrical conductivity and promise for transport applications. The magnetic easy axis lies in-plane, oriented 30° relative to the $y$-axis (Fig. 3b), while the $z$-direction represents the highest-energy configuration, lying 0.63 meV above the ground state.

Aligning the magnetization along the *x*- and *y*-axes reduces the total energy relative to the *z*-axis by 0.16 and 0.60 meV, corresponding to energy differences of ~0.01 and ~0.05 meV/atom, respectively. To circumvent the restriction imposed by the Mermin–Wagner theorem (*27*) and stabilize long-range magnetic order in 2D systems (*28*), uniaxial magnetic anisotropy is essential. The magnetic anisotropy energy (*29*) is given by

$$\mathrm{MAE} = \xi^2 \sum_{o,u} \sum_{\alpha,\beta} (1 - 2\delta_{\alpha\beta}) \frac{\left|\langle o^\alpha | L_x | u^\beta \rangle\right|^2 - \left|\langle o^\alpha | L_z | u^\beta \rangle\right|^2}{E_u^\beta - E_o^\alpha} \quad (1)$$

where the contributions originate from spin-orbit coupling (SOC) with coupling constant $\xi$, and $\delta_{\alpha\beta}$ denotes the Kronecker delta (it is 0 elsewhere and 1 when $\alpha = \beta$). $\alpha$ and $\beta$ stand for the spin orientations, *o* and *u* are the occupied and unoccupied states, respectively. Both Cr-3*d* and Sb-5*p* orbitals exhibit pronounced SOC interactions (Figs. 2c–h). The lowest-energy magnetization configuration lies near the *y*-direction, and the near-degeneracy with the ground state is reflected in similar intra-orbital SOC contributions (Figs. 2d–h). Rotating the easy axis toward the *x*-direction shifts the dominant positive contributions from Cr-($d_{z^2}$, $d_{xz}$) to Cr-($d_{xy}$, $d_{xz}$), ($d_{yz}$, $d_{x^2-y^2}$), and ($d_{yz}$, $d_{z^2}$), and from Sb-($p_z$, $p_x$) to Sb-($p_y$, $p_z$). The magnitudes of positive contributions diminishes while negative contributions arising from the same orbital channels Cr-($d_{xy}$, $d_{x^2-y^2}$) and Sb-($p_y$, $p_x$) increase slightly, narrowing the energy difference between the *x*- and *z*-axis configurations.

The two inner Cr atoms are coordinated by six Sb atoms distributed across the three atomic layers, whereas the outer Cr atoms are adjacent to vacuum on one side. This asymmetric environment induces differential charge accumulation (Fig. 3c), rendering the inner and outer Cr pairs inequivalent, with respective magnetic moments of ±2.6 and ±3.3 $\mu_B$. CrSb belongs to the spin space group of $P^{-1}m^{1}m^{-1}a^{\infty m}1$, within which the two sublattices of one atomic layer are related by operations such as $[C_2 \,||\, C_{2x}\tau_{1/2}]$ and $[C_2 \,||\, M_y\tau_{1/2}]$ (Fig. 3d). Crucially, along the out-of-plane direction, the AM sublattices residing in the inner and outer atomic layers cannot be connected by any symmetry operation, which is defining signature of FiAM. The momentum-dependent spin-resolved band dispersion obeys the symmetry constraints

$$[C_2 \,||\, C_{2x}\tau_{1/2}]\, E(k_x, k_y, \uparrow) = E(k_x, -k_y, \downarrow) \tag{2}$$

$$[C_2 \,||\, M_y\tau_{1/2}]\, E(k_x, k_y, \uparrow) = E(-k_x, k_y, \downarrow) \tag{3}$$

These account for the alternating spin-splitting observed along the M−Γ−M2 and M1−Γ−M paths (Figs. 3e and S1a). The spin-splitting energy near the Fermi level reaches 344 meV, which is larger than 2D AM $CoSe_2$ (213 meV) (*30*) and $VBr_4$ (150 meV) (*31*). By contrast, bands along M−X−Γ−Y−M path remain degenerate (Fig. S1b) owing to $E(k_x, 0, \uparrow) = E(k_x, 0, \downarrow)$ and $E(0, k_y, \uparrow) = E(0, k_y, \downarrow)$, as well as spin degeneracy at $k_x = m\pi$ and $k_y = n\pi$ enforced by the Bloch theorem (*30, 32*).

Notably, for FiAM CrSb, the band contributions from inner and outer atomic layers do not coincide in energy-momentum spaces (Figs. S2a-f): $OE_{\mathrm{inner}}(k_x, k_y, \uparrow/\downarrow) \neq E_{\mathrm{outer}}(k_x, k_y, \uparrow/\downarrow)$ with $O$ denoting any symmetry operation, while the two outer layers remain equivalent: $E_{\mathrm{outer1}}(k_x, k_y, \uparrow/\downarrow) = E_{\mathrm{outer2}}(k_x, k_y, \uparrow/\downarrow)$. The contribution from the inner atomic layer is relatively localized, with predominant weight concentrated around the Γ point near the Fermi level. Crucially, unlike conventional AM systems where reversing a subset of Néel vectors yields an AAM phase with entirely hidden spin splitting, the FiAM CrSb retains the overall spin splitting upon reversal of the Néel vectors of either the outer or inner atomic layer. This robustness against external perturbations is highly advantageous for spintronic transport. The spin texture near the Fermi level further corroborates the alternating spin splitting (Fig. 3f): around the Γ point, the upper portion of the bands closest to Fermi level exhibits predominantly spin-up character, while the lower portion (e.g., around $k_x \approx 0.2$ or $k_y \approx 0.2$) acquires a discernible spin-down component.

A magnetic transition temperature exceeding room temperature is essential for practical spintronic and data-storage applications (*33*). We therefore evaluate $T_{\mathrm{N}}$ of the three-atomic-layer CrSb (110) facet using a spin Heisenberg model with Hamiltonian

$$H = -[\sum_{<i,j>} J_{i,j} S_i S_j + \sum_{<k,l>} J_{k,l} S_k S_l + \sum_{<m,n>} J_{m,n} S_m S_n + \sum_{<o,p>} J_{o,p} S_o S_p + \sum_i A(S_i^z)^2] \tag{4}$$

where $J$, $S$, and $A$ denote exchange interaction, spin, and magnetic anisotropy, respectively. To extract numerical values of the four near-neighbor exchange parameters (Fig. 2a) from the energy differences among various magnetic configurations, we normalize the spin operators. The energy expressions for various magnetic states can be simplified as

$$E_{\mathrm{FM}} = -(E_0 + 2J_a + 8J_b + 6J_c + 8J_d + 6A) \tag{5}$$

$$E_{\mathrm{AFM1}} = -(E_0 + 2J_a + 8J_b - 6J_c - 8J_d + 6A) \tag{6}$$

$$E_{\mathrm{AFM2}} = -(E_0 + 2J_a - 8J_b - 6J_c + 8J_d + 6A) \tag{7}$$

$$E_{\mathrm{AFM3}} = -(E_0 - 2J_a + 0J_b - 6J_c + 0J_d + 6A) \tag{8}$$

$$E_{\mathrm{AFM4}} = -(E_0 + 0J_a - 4J_b + 2J_c - 4J_d + 6A) \tag{9}$$

in which $E_0$ is the total energy without magnetic coupling. Solving yields $J_a$ = 46.4, $J_b$ = 34.7, $J_c$ = -99.1, and $J_d$ = -0.5 meV. MC simulations employing the Metropolis algorithm give $T_N$ = 657 K (Fig. 3g), which approaches its bulk value of ~700 K (*23*) and is larger than those of many 2D altermagnets, such as FeS (405 K) and FeSe (190 K) (*34*), $Mn_2PSe$ (408 K) (*4*), and *t*-$Cr_2$[Pyc-$O_8$] (57 K) (*35*). This high $T_N$ confirms that long-range magnetic order can be persisted well above room temperature, underscoring the viability of 2D CrSb for pragmatic spintronic devices.

**Anomalous Transverse Phenomena**

Based on the magnetization easy axis of CrSb, we further examine the SOC-related anomalous transverse responses in altermagnets: the MOKE and the AHE. Both phenomena arise from broken time-reversal symmetry and manifest as transverse signals (Figs. 1b,c). MOKE is characterized by a relative rotation of the reflected light's polarization plane with reference to the incident light and an accompanying ellipticity change (Fig. 1b). The MOKE signal (*36, 37*) can be quantified by the complex Kerr angle $\Phi_K = \theta_K + i\varepsilon_K$, where $\theta_K$ and $\varepsilon_K$ denote the Kerr rotation angle and ellipticity, respectively. The substrate is $SiO_2$ with refractive index of $n$ = 1.546. For the CrSb (110) flakes (Fig. 3h), the absolute $\theta_K$ reaches 0.21 mrad at a photon energy of 3.70 eV. The $\varepsilon_K$ attains 0.37 mrad at 4.12 eV. The latter corresponds to a measurable variation in the minor-to-major axis ratio of the reflected polarization ellipse. These magneto-optical signatures (Fig. 1b) not only enable nondestructive characterization of magnetic orders but also highlight the promise of altermagnets for optical sensing technologies (*38, 39*).

The AHE in the three-atomic-layer CrSb (110) facet is evaluated via the intrinsic anomalous Hall conductivity (AHC) (*40*), which can be expressed as

$$\sigma_{xy} = \frac{e^2}{\hbar}\left(\frac{1}{2\pi}\right)^3 \int_{BZ} \Omega_z(k) d^3k \tag{10}$$

in which $e$, $\hbar = h/2\pi$, $k$, and $\Omega_z(k)$ stand for electron charge, reduced Planck's constant, crystal momentum, and Berry curvature, respectively. Within the Kudo formula framework, the Berry curvature can be described by $\Omega_z(k) = \sum_n f_n \Omega_n(k)$ with $f_n$ representing the Fermi-Dirac distribution (*41*). $\Omega_n(k)$ is given by

$$\Omega_n(k) = -2\mathrm{Im} \sum_{m \neq n} \frac{\hbar^2 \langle \Psi_{nk} | v_x | \Psi_{mk} \rangle \langle \Psi_{mk} | v_y | \Psi_{nk} \rangle}{\left(E_m(k) - E_n(k)\right)^2} \tag{11}$$

where $|\Psi_{mk}\rangle$ and $E_m$ are the Bloch eigenstate and eigenvalue, and the velocity operators along the $x$- and $y$-directions are $v_x$ and $v_y$, respectively. The transverse AHC of CrSb at the Fermi level is -11.6 S/cm, while the AHC reaches magnitudes on the order of hundreds of S/cm over the considered energy range (Fig. 3i). The dominant positive contribution to the Berry curvature arises from the vicinity of the Γ point and the region along the Γ-X line, whereas negative contributions are localized near the X, Y, and M points (Fig. 3j). Such anomalous transport behavior (Fig. 1c) positions FiAM CrSb as a promising candidate for Hall-effect devices and quantum coherent technologies (*31, 42*).

**Equilibrium Multiferroic Transport**

Beyond the altermagnetism above room temperature, MOKE, and AHE discussed above, we further evaluate the spin and charge transport performance of 2D CrSb within a multiferroic device architecture. A conventional spin valve (*43*) consists of two FM layers separated by a non-magnetic metallic spacer (Fig. 1f). Its magnetoresistance arises from the relative alignment of the FM electrodes, which governs the matching of spin-polarized transport channels. Analogously, a FE valve comprising two FE parts flanking a non-ferroic metallic spacer has been experimentally realized, yielding sizable electroresistance (*44*). Inspired by these paradigms, we propose an AM-FE valve: a non-ferroic metallic spacer is interposed between two FE parts, and the resulting trilayer is subsequently sandwiched between two AM electrodes to form a pentalayer structure (Fig. 1g). This architecture allows for both magnetoresistance and electroresistance effects, thereby facilitating multifunctional integration of spin and charge degrees of freedom.

The three-atomic-layer CrSb (110) facet is intrinsically centrosymmetric (*P* symmetry) and non-FE. Its orthogonal in-plane lattice constants are $a$ = 5.31 Å and $b$ = 7.18 Å. To assemble the proposed AM-FE valve, we select $Sc_2CO_2$, a well-established FE semiconductor (*45, 46*) from the MXene family obtainable via high-temperature dehydrogenation of $Sc_2C(OH)_2$ (*47*), as the FE component, and metallic Cu as the non-ferroic spacer. The calculated lattice constants of $Sc_2CO_2$ and Cu are 3.43 Å and 3.57 Å (Figs. S1c,d), respectively. $Sc_2CO_2$ exhibits a band gap of 1.86 eV (Fig. S1e), which is consistent with reported 1.83 eV (*48*), whereas Cu is metallic (Fig. 1f), with their respective high-symmetry points presented in Fig. S1g,h.

For transport along the *a*-direction of CrSb, $2\sqrt{3} \times 2$ and $1 \times 2$ supercells are adopted for $Sc_2CO_2$ and Cu, respectively. The lattice mismatches between the adjacent layers (7.18 Å of CrSb vs 6.86 Å of $Sc_2CO_2$ vs 7.14 Å of Cu) are within 4.5 %, which falls well inside the experimentally feasible range, as demonstrated for Fe/MgO/Fe (5%) (*49*) and graphene-contacted GaSe/$In_2Se_3$ (~7%) (*50*).

The CrSb/$Sc_2CO_2$/Cu/$Sc_2CO_2$/CrSb junctions are constructed (Figs. 4a,b), where the left portion is fixed: both the magnetic configuration of the left CrSb electrode and the FE polarization of the adjacent left $Sc_2CO_2$ are held invariant. In contrast, the Néel vector of the right CrSb electrode is switchable, enabling the relative alignment between the left and right electrodes ($M_P$ vs $M_{AP}$) that underpin TMR, which is a cornerstone for hard-disk read heads and magnetic random-access memories (*51-53*). Concurrently, the FE polarization of the right $Sc_2CO_2$ can be reversed, yielding parallel (P) and antiparallel (AP) FE configurations and providing a pathway to probe TER. It supports electrically controlled nonvolatile memory and neuromorphic computing (*54*).

The combination of magnetic and FE configurations yields four distinct states (Figs. 4a,b), for which the transmission coefficients in both energy and momentum spaces have been evaluated (Figs. 4c–n). The transmission coefficient is defined as $T = \sum_{k_{//}} T(\vec{k}_{//}) / N_k$, where $N_k$ denotes the number of *k*-points sampled in the 2D Brillouin zone, and the wave vector $\vec{k}_{//} = (k_x, k_y)$ lies in the plane perpendicular to the transport direction. Electron transmission is predominantly concentrated between the Γ-X and Y-M lines, running parallel to these segments (Figs. 4g–n).

In equilibrium conditions, the spin filtering efficiency ($\eta$) reaches -85% and -87% for $M_P$-P and $M_{AP}$-AP states, respectively (Table 1). These arise from that spin-down electrons tunnel far more readily, dominating the overall transmission. The negative high values reflect high spin fidelity during transport (*55*). In contrast, spin filtering is weak for the $M_P$-AP and $M_{AP}$-P states. For the $M_{AP}$ configuration, the spin splitting of the right electrode is inverted relative to the left electrode, resulting in substantial channel mismatch and a transition from the low-resistance ($M_P$) state to high-resistance ($M_{AP}$) state (Fig. 1d). The TMR ratios reach 1,677% and 470% for the $M_P$-P vs $M_{AP}$-P and $M_P$-AP vs $M_{AP}$-AP, respectively. This stems from the electronic structures matching in $M_P$, which facilitates electron tunneling and yielding larger transmission coefficients than those in $M_{AP}$ cases. The value of 1,677% is higher than the equilibrium values of 1,056% across $CrSb/In_2Se_3/Fe_3GaTe_2$ (*17*) and 574% across $Ag/V_2Te_2O/BiOCl/V_2Te_2O/Ag$ (*56*). Regarding electrical transport, the P configuration aligns the FE polarizations of the two $Sc_2CO_2$ in the same direction, generating cooperative built-in electric fields that enhance quantum tunneling. Conversely, the AP configuration induces opposing polarizations and conflicting electric fields, substantially impeding electron tunneling and giving rise to a high-resistance state. Consistently, electronic transmission is stronger in P than in AP (Table 1). The TER ratios attain 4,764% for $M_P$-P vs $M_P$-AP and 1,461% for $M_{AP}$-P and $M_{AP}$-AP. These values are higher than 707% across $CrSb/In_2Se_3/Fe_3GaTe_2$ (*17*) and 821% across $Mn_3GaN/BaTiO_3/Mn_3ZnN$ (*57*).

**Non-Equilibrium Multi-Field Altermagnetic-Ferroelectric Valve**

We further extend the transport properties to non-equilibrium states by incorporating a finite bias voltage (Figs. 5a–f). For $M_P$-P, the spin filtering effect remains pronounced across the entire applied bias range, indicating that this state serves as a near-perfect filter for spin-up electrons (Fig. 5a). In the other configurations (Figs. 5b, d, and e), the spin filtering efficiency can also reach appreciable levels upon bias variation. The TMR attains $3.1\times10^4$% for $M_P$-P vs $M_{AP}$-P at a -0.06 V, and $1.2\times10^4$% for $M_P$-AP vs $M_{AP}$-AP at a -0.08 V. The TER is also significantly enhanced under bias, reaching $1.1\times10^4$% of $M_P$-P vs $M_P$-AP within the bias from -0.04 to -0.02 V. In addition, NDR behavior, which is characterized by a decrease in current with increasing bias voltage, emerges in several configurations: in $M_P$-AP within the voltage range from 0.04 (0.06) to 0.08 V for the spin-up (spin-down) channel (Fig. 5b); in $M_{AP}$-P from 0.08 to 0.10 V for the spin-down channel (Fig. 5d); and in $M_{AP}$-AP for the spin-up channel from 0.06 to 0.08 V (Fig. 5e). NDR originates from resonant tunneling: when incident electrons energies align with quantized levels in the quantum well, transmission is maximized. Increasing the bias voltage disrupts this resonance condition, causing a precipitous drop in tunneling current. The FE materials act as tunable barriers: their polarization orientation directly controls the energetic positions of resonant levels, thereby enabling electrical gating of the NDR onset. These findings underscore the superior non-equilibrium performance of the proposed AM-FE valve and highlight its potential for multi-level cell functionality in next-generation miniaturized memory devices.

Applying a bias voltage across the AM-FE valve ensures operational stability and control, harvesting ambient thermal or optical energy to drive device operation offers a compelling alternative strategy. Such self-sustained modalities reduce reliance on external circuitry, lower system cost, and enable environment-friendly, extended-lifetime functionality (*58, 59*). We therefore examine thermally driven spin transport by imposing a fixed temperature difference of 20 K between the left ($T_L$) and right ($T_R$) electrodes, while varying $T_L$ from 100 K to 600 K (Figs. 5g–l). The thermal spin current is governed by two principal factors: the Fermi-Dirac distribution difference between the electrodes, and the energy-dependent transmission coefficient integrated over the window where this difference is non-negligible (*42*). Appreciable spin filtering efficiencies are achieved: -90% for $M_P$-P at 540 K (Fig. 5g), 82% to 90% for $M_P$-AP from 160 to 360 K (Fig. 5h), -82% to -84% for $M_{AP}$-P at 100–140 K (Fig. 5j), and -68% for $M_{AP}$-AP at 600 K (Fig. 5k). The thermal TMR reaches $2.2\times10^4$% for $M_P$-P vs $M_{AP}$-P at $T_L$ = 520 K and $4.2\times10^4$% for $M_P$-AP vs $M_{AP}$-AP at $T_L$ = 300 K (Fig. 5i). The thermal TER attains $7.0\times10^3$% for $M_P$-P vs $M_P$-AP and $3.3\times10^4$% for $M_{AP}$-P vs $M_{AP}$-AP at $T_L$ = 100 K (Fig. 5l). These results establish that the AM-FE valve, operating solely under a temperature differential without an external power source, can deliver well-defined resistive signals and is thus attractive for low-power applications including wearable electronics and biomedical implants.

Extending the self-powered paradigm, we examine spin and charge transport under optical illumination (Figs. 6a–j). Photon energies of 1, 2, and 3 eV are selected as representative values, and the influence of light polarization is probed by varying the polarization angle ($\theta$) relative to the transport direction: $\theta$ = 0° (90°) corresponds to the light polarization parallel (perpendicular) to the transport direction. For $M_P$-P and $M_P$-AP configurations under 3 eV illumination (Figs. 6a,c), the photocurrents in both spin channels exhibit a distinct second-order response, indicating the photogalvanic effect. It is superimposed on a conventional photovoltaic background. This effect arises from the built-in electric field generated by the broken $P$ symmetry in the FE $Sc_2CO_2$ and enables bias-free electron transport.

Near-perfect spin filtering efficiencies are achieved under illumination: 95% for $M_P$-P at photon energy ($E_P$) = 1 eV with $\theta$ = 105° (Figs. 6a); -81% for $M_{AP}$-P at $E_P$ = 1 eV with $\theta$ = 0° and 180° (Fig. 6b); -88%/-97%/-91% for $M_P$-AP at $E_P$ = 1/2/3 eV with $\theta$ = 120°/45°/150° (Fig. 6c); and 90%/96% for $M_{AP}$-AP at $E_P$ = 2/3 eV with $\theta$ = 120°/60° (Fig. 6d). These values demonstrate selective generation of spin-polarized photocurrents via optical approach. The photocurrents for $M_{AP}$ configurations (Figs. 6b,d) are comparatively smaller, because the mismatch in the spin splitting of the left and right electrodes impedes electron tunneling. Under optical excitation, both TMR and TER can reach giant magnitudes. The optical TMR attains $5.1\times10^3$% for $M_P$-AP vs $M_{AP}$-AP with $\theta$ = 30°, $9.0\times10^4$% for $M_P$-P vs $M_{AP}$-P with $\theta$ = 30°, and $1.6\times10^4$% for $M_P$-AP vs $M_{AP}$-AP with $\theta$ = 120° at $E_P$ = 1, 2, and 3 eV, respectively (Figs. 6e–g). A giant optical TER of $5.6\times10^4$% is obtained at $E_P$ = 2 eV with $\theta$ = 165° (Fig. 6i). Large values are also achievable at $E_P$ = 1 and 3 eV: 804% for $M_P$-P vs $M_P$-AP at $E_P$ = 1 eV with $\theta$ = 60° (Fig. 6h) and $2.2\times10^3$% for $M_{AP}$-P vs $M_{AP}$-AP at $E_P$ = 3 eV with 120° (Fig. 6j). These position the AM-FE valve as a promising platform for quantum communication and nano-sensing applications.

Moreover, we quantify the optical switching contrast via ER = $I_{max}/I_{min}$, which characterizes the ability to distinguish between "on" and "off" optical signal states. For comparison, the ratio of the photocurrents generated under parallel and perpendicular light polarization is also evaluated (Tables 2 and 3). The spin-down ER reaches 283.8 for the $M_{AP}$-AP configuration at $E_P$ =3 eV, which is larger than those reported for bent CrSBr (123) (*60*) and iodine-vacancy perovskite $CsCu_2I_3$ (69.7) (*61*). ER values on the order of several tens are also widely achieved in CrSb/$Sc_2CO_2$/Cu/$Sc_2CO_2$/CrSb. A large extinction ratio enhances the signal-to-noise ratio, reduces the bit error rate, and minimizes system-level power dissipation.

## DISCUSSION

In summary, we have proposed ferrialtermagnetism as a distinct subclass of AM order, wherein symmetry-inequivalent AM sublattices with nonidentical Néel vectors preclude the mutual cancellation of alternating spin splitting. This concept is concretely realized in three-atomic-layer CrSb (110) flakes, where the asymmetric coordination environments of inner and outer Cr layers naturally break the symmetry connectivity required for conventional or hidden altermagnetism. The FiAM state exhibits alternating spin splitting of 344 meV and a high $T_N$ of 657 K. The MOKE and AHE are observed. Furthermore, we have designed an AM-FE valve by integrating CrSb electrodes with $Sc_2CO_2$ and Cu spacer. This pentalayer architecture yield giant equilibrium TMR of 1,677% and TER of 4,764%. Extending into non-equilibrium regimes, this device exhibits bias TMR/TER of $3.1\times10^4$%/$1.1\times10^4$%, thermal TMR/TER of $4.2\times10^4$%/$3.3\times10^4$%, and optical TMR/TER of $9.0\times10^4$%/$5.6\times10^4$%. High spin-filtering efficiencies around 90% can be achieved under both equilibrium and non-equilibrium conditions. Pronounced NDR and photogalvanic effects, and high ER of 283.8 are observed. These findings demonstrate that the AM multiferroic integration and multi-field control over both spin and charge transport, motivating AM family extension and advancing self-powered spintronic and electronic miniaturized devices.

## MATERIALS AND METHODS

First-principles calculations were performed within the Vienna Ab initio Simulation Package (VASP) based on the DFT (*62*). The generalized gradient approximation (GGA) with the Perdew–Burke–Ernzerhof (PBE) exchange–correlation functional was utilized (*63*). The PBE results were in good agreement with experimental results including angle-resolved photoemission spectroscopy (ARPES) measurements (*64, 65*). A cutoff energy 600 eV was adopted, with convergence criteria of $10^{-7}$ eV for energy and 0.001 eV/Å for force. The Γ-centered 8 × 6 × 1 Monkhorst-Pack grid was used for three-atomic-layer CrSb. A vacuum space of 15 Å was employed for eliminating the interlayer interactions from the periodic structures. MC simulations were performed using the MCSOLVER code (*66*) with a 60 × 60 × 1 supercell. The P4VASP (*67*) and VASPKIT (*68*) packages were used for pre- and post-processing the data from VASP. The VASPBERRY code was utilized for calculating Berry curvature from the wave function (*69*), and the Wannier90 codes were used to calculate the AHC (*70*). The full lattice and atomic relaxations were performed.

Combing DFT with the NEGF approach, the spin transport properties were calculated in the QuantumWise Atomistix ToolKit (ATK) package (*71*). The 1 × 6 × 150 and 50 × 50 meshes were utilized for self-consistent and transmission calculations, respectively. The cutoff energy was 150 hartree. For equilibrium conditions, the spin filtering efficiency can be defined as $\eta = (T_{\uparrow} - T_{\downarrow}) / (T_{\uparrow} + T_{\downarrow}) \times 100\%$, in which the $T_{\uparrow}$ and $T_{\downarrow}$ are the transmission coefficients for spin-up and spin-down channels, respectively. The TMR is obtained by $\mathrm{TMR} = (T_{\mathrm{M_P}} - T_{\mathrm{M_{AP}}}) / T_{\mathrm{M_{AP}}} \times 100\%$, where the $T_{\mathrm{M_P}}$ and $T_{\mathrm{M_{AP}}}$ are the total transmission coefficients for the parallel and antiparallel magnetizations of CrSb, respectively. Similarly, TER is attained by $\mathrm{TER} = (T_{\mathrm{P}} - T_{\mathrm{AP}}) / T_{\mathrm{AP}} \times 100\%$, with the $T_{\mathrm{P}}$ and $T_{\mathrm{AP}}$ representing transmission coefficients for the opposite FE polarization directions of the $Sc_2CO_2$. Based on the Landauer–Büttiker formula (*72, 73*), the spin-dependent current of the device at a bias can be written as $I_{\uparrow(\downarrow)} = \frac{e}{h}\int \left\{T_{\uparrow(\downarrow)}(E,V_b)\left[f_{\mathrm{L}}(E-\mu_{\mathrm{L}}) - f_{\mathrm{R}}(E-\mu_{\mathrm{R}})\right]\right\}\mathrm{d}E$, where the electrochemical potential of the left/right electrode at bias $V_b$ is $\mu_{\mathrm{L/R}} = E_{\mathrm{F}} \pm eV_b / 2$. The $f_{\mathrm{L/R}}$ is the Fermi-Dirac distributions for the left/right electrodes. Similarly, the thermal current can be obtained by $I_{\uparrow(\downarrow)} = \frac{e}{h}\int \left\{T_{\uparrow(\downarrow)}(E)\left[f_{\mathrm{L}}(E,T_{\mathrm{L}}) - f_{\mathrm{R}}(E,T_{\mathrm{R}})\right]\right\}\mathrm{d}E$, where $f_{\mathrm{L/R}}(E,T_{\mathrm{L}}) = 1/[\exp(E/kT)+1]$. The photocurrent can be obtained by $I = \frac{ie}{h}\int Tr\left\{\Gamma_{\mathrm{L}}[G^{<(ph)} + f_{\mathrm{L}}(E)(G^{>(ph)} - G^{<(ph)})]\right\}\mathrm{d}E$, with $G^{>(ph)}$ and $G^{<(ph)}$ representing the greater and lesser Green's function of the electron-photon interaction system (*74*). For transport under external field (bias voltage, temperature gradient, and linearly polarized light), spin-polarized currents are yielded. The spin filtering efficiency can be written as $\eta = (|I_{\uparrow}| - |I_{\downarrow}|) / (|I_{\uparrow}| + |I_{\downarrow}|) \times 100\%$, with $I_{\uparrow}$ and $I_{\downarrow}$ denoting the currents in spin-up and spin-down channels, respectively. The TMR can be calculated by $\mathrm{TMR} = (|I_{\mathrm{M_P}}| - |I_{\mathrm{M_{AP}}}|) / |I_{\mathrm{M_{AP}}}| \times 100\%$, where the $I_{\mathrm{M_P}}$ and $I_{\mathrm{M_{AP}}}$ are the total charge currents for the parallel and antiparallel magnetizations of CrSb, respectively. The TER is calculated by $\mathrm{TER} = (|I_{\mathrm{P}}| - |I_{\mathrm{AP}}|) / |I_{\mathrm{AP}}| \times 100\%$, in which the $I_{\mathrm{P}}$ and $I_{\mathrm{AP}}$ representing charge currents for the opposite FE polarization directions of the $Sc_2CO_2$.

## Supplementary Materials

### This PDF file includes:

Figs. S1 to S2

**Acknowledgments:** L. Z. acknowledges support from the Doctoral Student Program of the Young S&T Talents Cultivation Project, CAST. The authors acknowledge Beijing PARATERA Technology Co., LTD for providing high-performance and AI computing resources for contributing to the research results reported within this paper. URL: http://www.cloud.paratera.com. **Funding:** G.D. acknowledges support from the National Natural Science Foundation of China (Grant No. 12374002). G.G. acknowledges support from the National Natural Science Foundation of China (Grant No. 12174127). **Author contributions:** L.Z. and G.G. conceived the idea and guided the research. L.Z. performed the theoretical simulations and wrote the paper. L.Z., X.C., H.L., J.D., F.Z., Y.Y., G.D., and G.G. contributed to data analysis. G.D. and G.G. supervised the research. All authors provided feedback on the manuscript and discussed the findings. **Competing interests:** the authors declare that they have no competing interests. **Data and materials availability:** All data needed to evaluate the conclusions in the paper are present in the paper and/or the Supplementary Materials.

## Figures and Tables

**Table 1. Spin-resolved transmission.** Spin-dependent electron transmission, TMR, TER, and $\eta$ across $CrSb/Sc_2CO_2/Cu/Sc_2CO_2/CrSb$.

| | $M_P$ | | | | $M_{AP}$ | | | | |
|---|---|---|---|---|---|---|---|---|---|
| | $T_\uparrow$ | $T_\downarrow$ | $T_{tot}$ | $\eta$ (%) | $T_\uparrow$ | $T_\downarrow$ | $T_{tot}$ | $\eta$ (%) | TMR (%) |
| P | $1.58 \times 10^{-4}$ | $1.99 \times 10^{-3}$ | $2.15 \times 10^{-3}$ | -85 | $5.03 \times 10^{-5}$ | $7.10 \times 10^{-5}$ | $1.21 \times 10^{-4}$ | -17 | 1,677 |
| AP | $2.81 \times 10^{-5}$ | $1.61 \times 10^{-5}$ | $4.42 \times 10^{-5}$ | 27 | $5.23 \times 10^{-7}$ | $7.23 \times 10^{-6}$ | $7.75 \times 10^{-6}$ | -87 | 470 |
| TER (%) | 4,764 | | | | 1,461 | | | | |

**Table 2. Spin-up optical switching.** Current ratio for $\theta$ = 90° vs 0°, 0° vs 90°, and maximum vs minimum in spin-up channel.

| Spin up | $\lvert I_\perp/I_{/\!/}\rvert$ | | | $\lvert I_{/\!/}/I_\perp\rvert$ | | | $\lvert I_{max}/I_{min}\rvert$ | | |
|---|---|---|---|---|---|---|---|---|---|
| $E_P$ (eV) | 1 | 2 | 3 | 1 | 2 | 3 | 1 | 2 | 3 |
| $M_P$-P | 1.2 | 3.6 | 0.3 | 0.9 | 0.3 | 3.0 | 3.9 | 24.5 | 3.0 |
| $M_{AP}$-P | 7.7 | 0.1 | 0.2 | 0.1 | 8.9 | 6.2 | 8.8 | 67.1 | 6.2 |
| $M_P$-AP | 1.3 | 1.0 | 1.3 | 0.8 | 1.0 | 0.7 | 15.8 | 11.3 | 9.6 |
| $M_{AP}$-AP | 1.8 | 1.2 | 0.4 | 0.6 | 0.9 | 2.6 | 4.5 | 1.8 | 2.8 |

**Table 3. Spin-down optical switching.** Current ratio for $\theta$ = 90° vs 0°, 0° vs 90°, and maximum vs minimum in spin-down channel.

| Spin down | $\lvert I_\perp/I_{/\!/}\rvert$ | | | $\lvert I_{/\!/}/I_\perp\rvert$ | | | $\lvert I_{max}/I_{min}\rvert$ | | |
|---|---|---|---|---|---|---|---|---|---|
| $E_P$ (eV) | 1 | 2 | 3 | 1 | 2 | 3 | 1 | 2 | 3 |
| $M_P$-P | 0.6 | 1.5 | 0.7 | 1.7 | 0.7 | 1.4 | 36.7 | 8.3 | 1.5 |
| $M_{AP}$-P | 0.5 | 0.7 | 0.9 | 1.9 | 1.5 | 1.1 | 33.2 | 2.0 | 1.4 |
| $M_P$-AP | 5.0 | 0.9 | 1.2 | 0.2 | 1.1 | 0.8 | 5.0 | 28.9 | 11.2 |
| $M_{AP}$-AP | 0.1 | 0.3 | 0.3 | 12.1 | 2.9 | 8.2 | 20.0 | 44.1 | 283.8 |

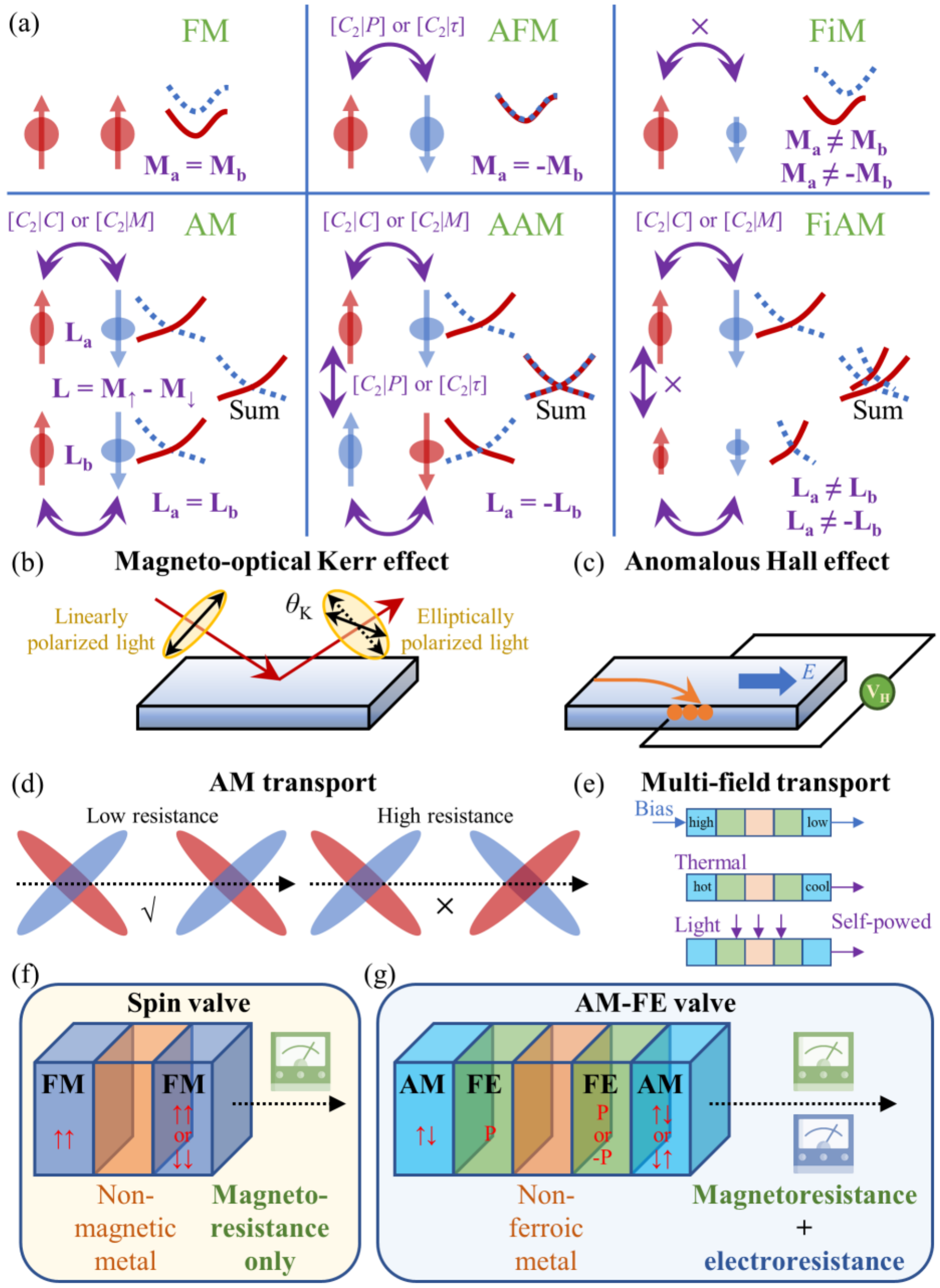

**Fig. 1. Symmetries of magnetic systems, and AM applications.** The symmetry connection, magnetization, and spin splitting of collinear FM, AFM, FiM, AM, AAM, and FiAM systems (a). Schematic diagrams of the MOKE (b) and the AHE (c). Schematic low and high resistances in AM transport (d). Multi-field transport using bias, thermal, and light (e). Schematics of the spin valve with magnetoresistance (f) and the AM-FE valve with magnetoresistance and electroresistance (g).

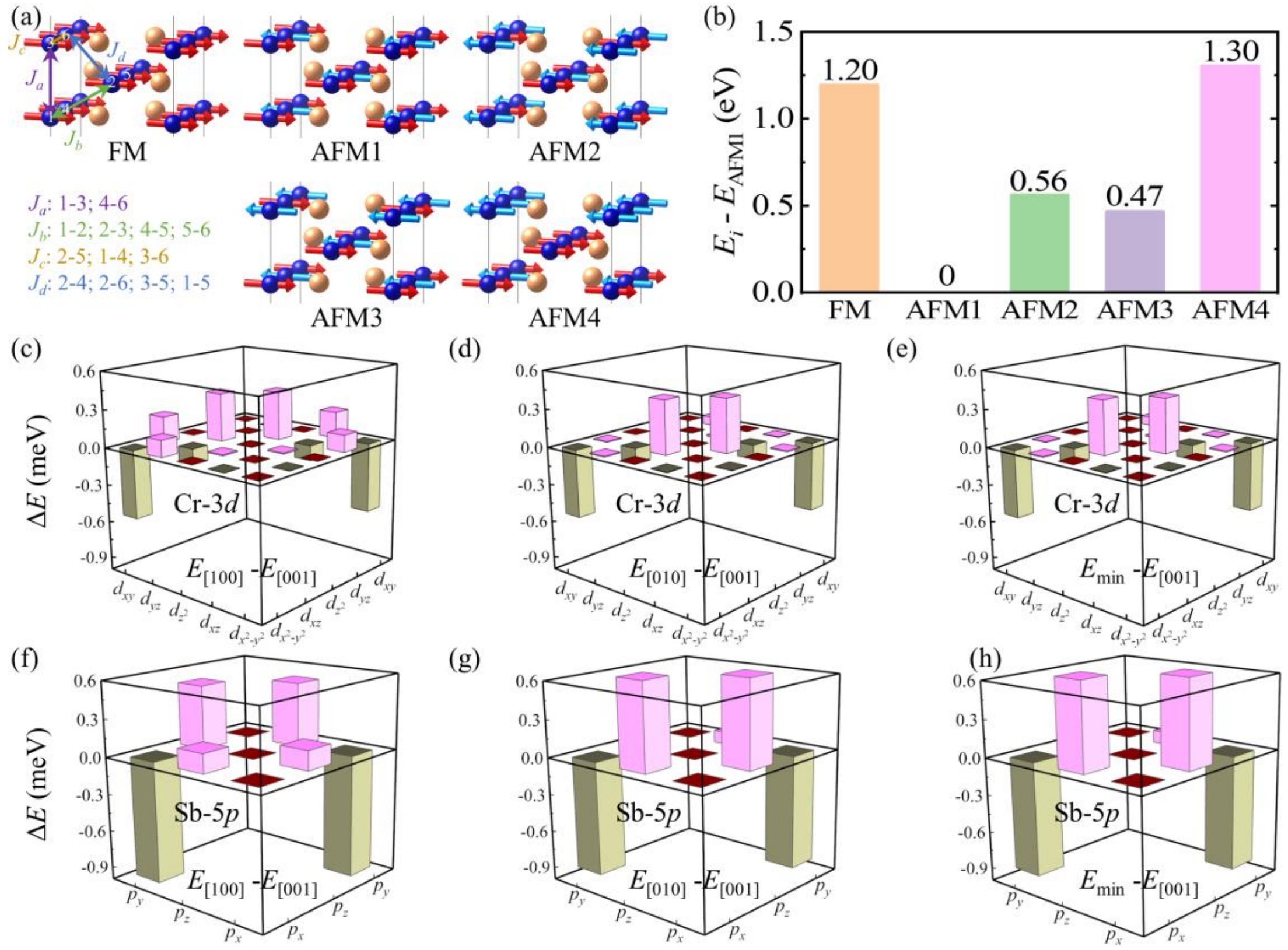

**Fig. 2. Magnetic states and anisotropy.** Various magnetic configurations in CrSb with considered exchange interactions (a). The energy difference $\Delta E = E_i - E_{\mathrm{AFM1}}$ (b). Cr-3*d*-contributed (c-e), Sb-5*p*-contributed (f-h) anisotropy energy.

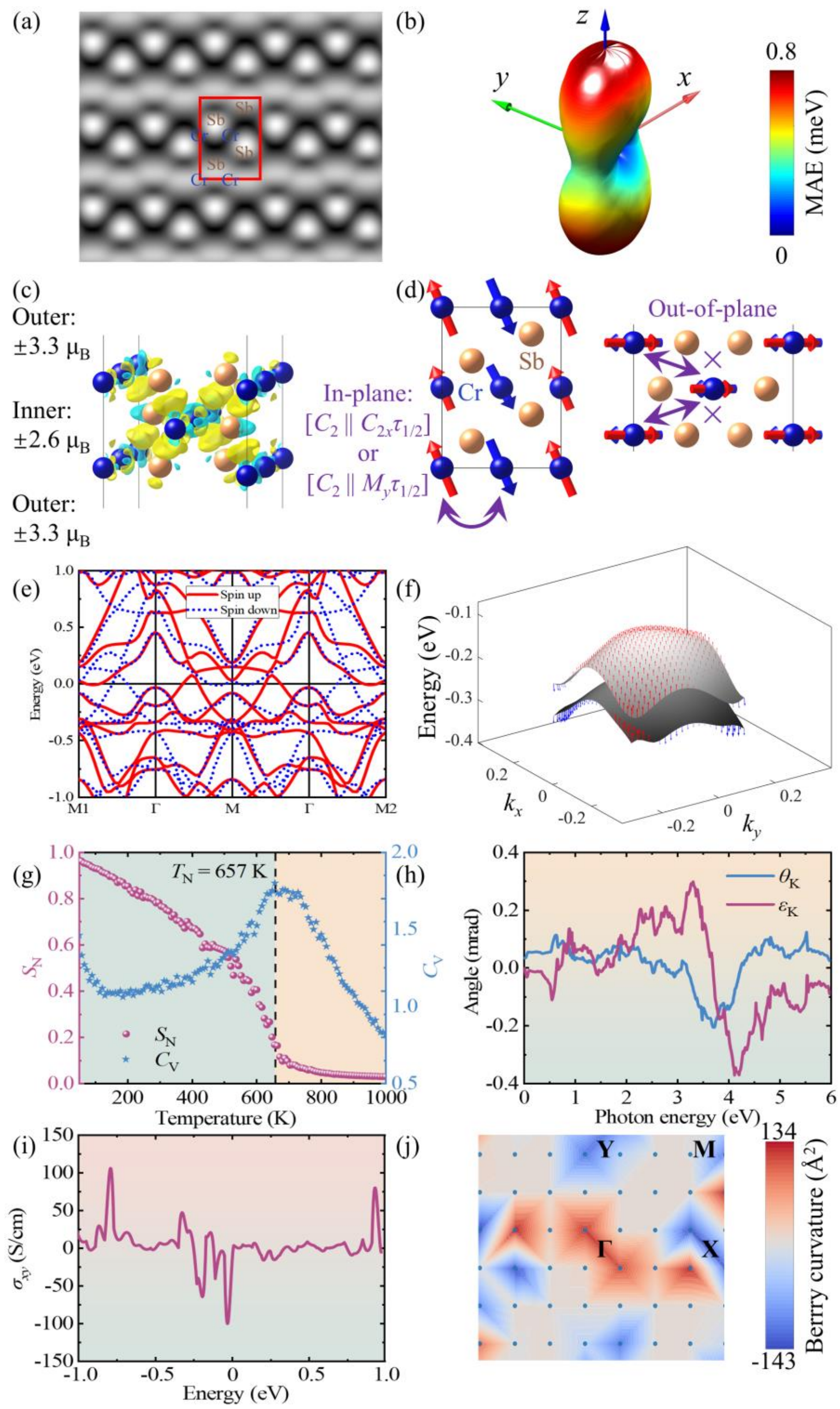

**Fig. 3. Theoretical characterization, $T_N$, MOKE, and AHC.** Simulated STM image of CrSb, where the red box is a unit cell (a). MAE in 3D space (b). Charge density difference for CrSb with an isosurface of 0.005 e/bohr$^3$ and Cr atomic magnetic moments (c). Structure of AM CrSb with connections between two sub-lattices (d). Alternating spin splitting (e). Momentum-dependent spin texture around the Fermi level (f). The variations of the normalized spin (red) and specific heat capacity (blue) with the temperature (g). The $\theta_K$ and $\varepsilon_K$ with varying photon energies (h). Energy-dependent AHC (i) and momentum-dependent Berry curvature (j).

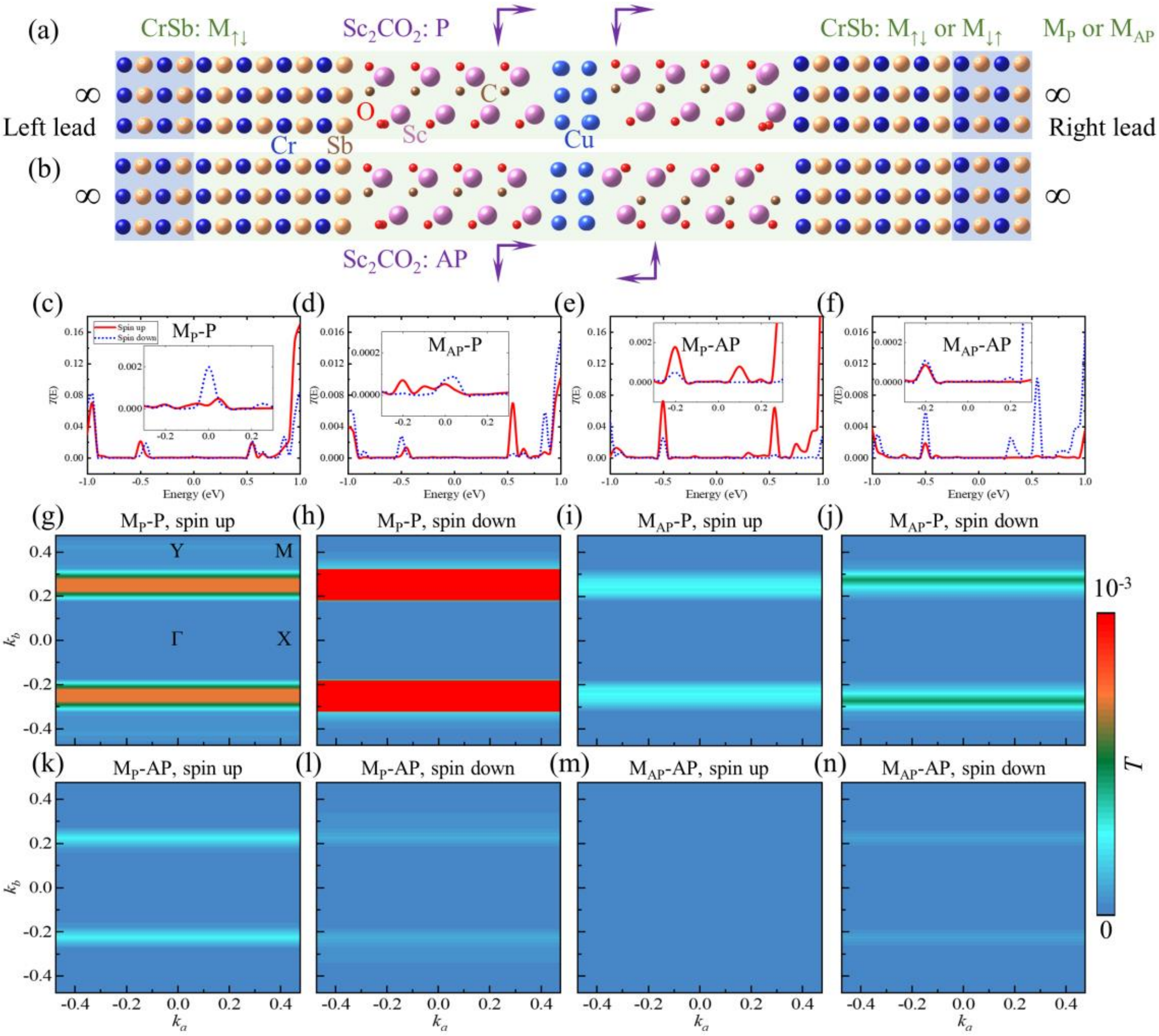

**Fig. 4. Transport performance.** Structural models of $CrSb/Sc_2CO_2/Cu/ScCO_2/CrSb$ with opposite FE polarization directions (a,b). Transmission coefficients as a function of energy with $M_P$-P (c), $M_{AP}$-P (d), $M_P$-AP (e), and $M_{AP}$-AP (f). Momentum-resolved transmission spectra (g-n). $M_P$ and $M_{AP}$ denote the same and opposite magnetization between the left and right CrSb electrodes, P and AP are the parallel and antiparallel FE polarization directions between the left and right $Sc_2CO_2$, respectively.

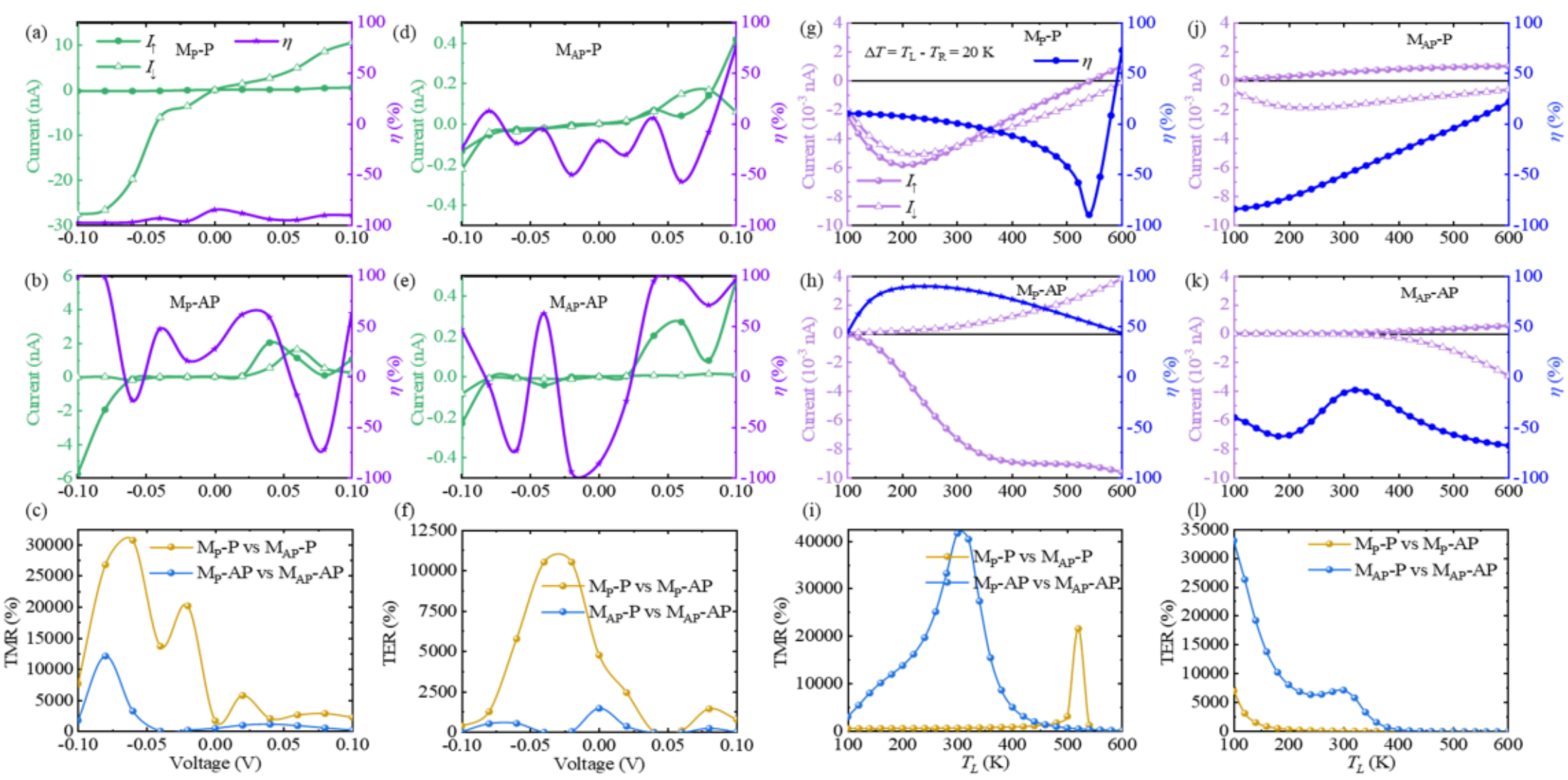

**Fig. 5. Bias and thermal transport.** Spin-dependent current, and $\eta$, TMR, and TER across CrSb/$Sc_2CO_2$/Cu/$Sc_2CO_2$/CrSb under bias voltage (a-f) and temperature gradient (g-l).

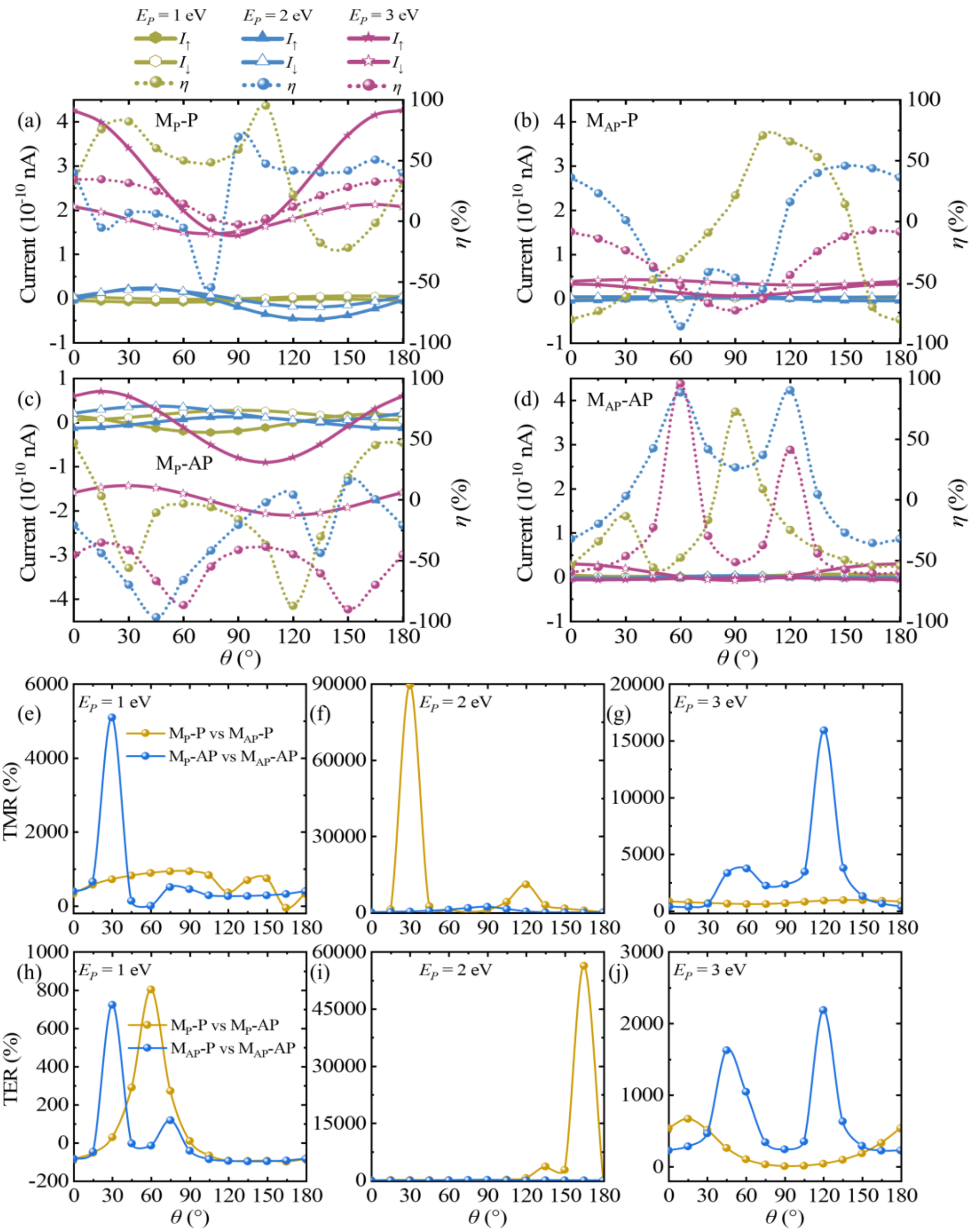

**Fig. 6. Photoelectric transport.** Spin-dependent current, and $\eta$, TMR, and TER across CrSb/$Sc_2CO_2$/Cu/$Sc_2CO_2$/CrSb with various $\theta$ at $E_P$ = 1, 2, and 3 eV.